\documentclass{nws}

\pdfoutput=1

\usepackage[colorlinks, breaklinks,hyperfootnotes = false,citecolor = red,naturalnames]{hyperref} 
 \DeclareMathAlphabet{\mathcal}{OMS}{cmsy}{m}{n}
\usepackage{graphicx}

\newcommand{\mS}[0]{$\mathcal{S}$}
\newcommand{\mI}[0]{$\mathcal{I}$}
\newcommand{\mM}[0]{$\mathcal{M}$}
\newcommand{\E}[0]{$\boldsymbol{\varepsilon}$}

\newcommand{\eref}[1]{(\ref{#1})}

\begin{document}

\title[Quantifying edge-to-edge relations in complex networks]{Structure of complex networks: Quantifying edge-to-edge relations by failure-induced flow redistribution}

 \author[M.T. Schaub et al.]
        {Michael T. Schaub\\
         Department of Mathematics, Imperial College London, South Kensington campus, London SW7 2AZ, U.K.\\
	  J\"org Lehmann\\
	  ABB Switzerland Ltd, Corporate Research, CH-5405 Baden-D\"attwil, Switzerland\\
	 Sophia N. Yaliraki\\
         Department of Chemistry, Imperial College London, South Kensington campus, London SW7 2AZ, U.K.\\
	 Mauricio Barahona\\
         Department of Mathematics, Imperial College London, South Kensington campus, London SW7 2AZ, U.K.
	\email{michael.schaub09@imperial.ac.uk, s.yaliraki@imperial.ac.uk, joerg.lehmann@ch.abb.com, m.barahona@imperial.ac.uk}}

 \jdate{July 2013}
\pubyear{2013}

\maketitle

\begin{abstract}
The analysis of complex networks has so far revolved mainly around the role of nodes and communities of nodes.  However, the dynamics of interconnected systems is often focalised on edge processes, and a dual edge-centric perspective can often prove more natural.
Here we present graph-theoretical measures to quantify edge-to-edge relations inspired by the notion of flow redistribution induced by edge failures.
Our measures, which are related to the pseudo-inverse of the Laplacian of the network, are global and reveal the dynamical interplay between the edges of a network, including potentially \textit{non-local} interactions.
Our framework also allows us to define the embeddedness of an edge, a measure of how strongly an edge features in the weighted cuts of the network.
We showcase the general applicability of our edge-centric framework through analyses of the Iberian Power grid, traffic flow in road networks, and the  \textit{C. elegans} neuronal network.
\end{abstract}

\section{Introduction}
The use of network formulations for the analysis of complex systems has attracted tremendous interest over the last years.
Network-centric approaches, in which the entities (agents, particles) of a system are represented as nodes in a graph and their interactions are denoted by (weighted, directed, multiplex) edges between nodes, have been successfully employed to model biological, technical and social systems~\cite{Albert2002,Boccaletti2006,Arenas2008a}.
The trend towards this network perspective has been facilitated by the increased availability of large relational datasets and growing computational resources.
Inevitably, this data-driven approach has led to the generation of large, highly complex networks.
However, such networks have limited explicative power and further analysis is usually needed to extract relevant representations from system interactions.
In this context, \textit{community detection} aims at obtaining coarse-grained, simplified descriptions of a network based on groups of nodes (i.e., communities) which can provide insight about the structure and function of the overall system~\cite{Schaeffer2007,Fortunato2010}.

Thus far, the majority of research on complex networks has focused on nodes, their roles, and their groupings into meaningful communities.
However, in a number of scenarios it is the dynamics on the edges and their interplay that defines the behaviour of the system.
Consider the generic case in which edges carry a flow (signal, data, mass, energy, etc) and where fluctuations or total/partial failures on edges can occur or be induced.
If the direct path between nodes A and B is blocked and only a fraction of the original flow can be transmitted, this blockade can cascade through the network affecting the flow on other links.
In this case, \textit{edge variables} and their mutual influences constitute the object of interest in the modeling.
The duality between edge and node based descriptions is at the heart of applications in circuit theory (even with nonlinear elements~\cite{Barahona1997,Barahona1998}), computational mechanics, estimation theory, as well as Systems Engineering and primal/dual problems  in optimisation theory.
In all these cases, an equivalent edge formulation can be exploited to highlight the relevance of processes focalised on the edges (or the cycles) rather than on the nodes of the network~\cite{Strang1986}. (See the Appendix for further connections to classic work in these areas.) 
However, such an edge-centric analysis has not been a focus in the recent literature of complex networks, in which graph-theoretical notions based on edges, such as the \textit{line graph}~\cite{Harary1960, Godsil2001} that records the immediate adjacency of edges, have only been used to investigate overlapping node communities in networks~\cite{Evans2009,Ahn2010}.

In the following, we introduce such an edge-centric framework.
Specifically, we derive an \textit{edge-to-edge matrix} based on the redistribution of linear flow under perturbations to the network and rewrite this matrix in terms of global graph-theoretical measures that quantify the specific architecture of edge-to-edge influences and the likelihood that each edge is critical to flow redistribution in the network.
Our derivation relates these notions explicitly to generic algebraic graph properties.
The analysis of this edge-to-edge matrix allows us to uncover potentially long-range relations between edges and can reveal non-local features in the organisation of complex networks.
We exemplify the general applicability of our measures with analyses of the Iberian Power grid, traffic flow in road networks, and the \textit{C. elegans} neuronal network. 

\subsection*{Notation}
We consider connected, weighted, undirected graphs with $N$ nodes (or vertices) and $E$ edges (or links).
Each edge $e$ is endowed with an arbitrary (but fixed) `reference' direction from the tail node $t(e)$ to its head $h(e)$.
Note that the graph is still \textit{undirected}: the flow is allowed to pass in both directions along each edge and the reference direction merely specifies the sign of the flow on the edge.
Each edge $e$ is associated with a $N \times 1$ incidence vector $\mathbf{b}_e$  with entries $[b_e]_{h(e)} = -1$, $[b_e]_{t(e)} = 1$ and zero otherwise.
Note that other authors use the opposite sign convention for $\mathbf{b}_e$.
The node-to-edge incidence matrix is then written as:
\begin{equation*}
  B_{N \times E} =[ \mathbf{b}_1 \cdots \mathbf{b}_E ].
\end{equation*}
Each edge $e$ has an associated (positive) weight or conductance $g_e$,  which we compile into a diagonal matrix
\begin{equation}
G_{E \times E} = \mathrm{diag}(g_e).
\end{equation}
The (weighted) graph Laplacian or Kirchhoff conductance matrix $L$ is then:
\begin{equation}\label{eq:laplacian}
  L_{N \times N} = \sum_{e=1}^E g_e \mathbf{b}_e \mathbf{b}_e^T = BGB^T.
\end{equation}
For connected, undirected graphs, $L$ is symmetric positive semidefinite, with a simple zero eigenvalue and corresponding eigenvector $\textbf{1}$, the vector of ones~\cite{Mohar1992,Mohar1997}.
In the following, node variables are denoted by capital letters while small letters are reserved for edge quantities.

\section{Edge-to-Edge relationships based on flow redistribution}

\subsection{The flow-redistribution matrix $K$}
As a means to make our formulation of linear flows more concrete, we introduce our framework through the canonical example of electrical resistor networks~\cite{Guattery1998, Strang1986} and its well-known connection with random walks~\cite{Doyle1984}.
Indeed, electrical resistor networks are not only relevant for electrical engineering applications, but can also be seen as archetypal models for linear processes of interest in various biological applications, e.g., vision~\cite{Poggio1985,Hutchinson1988}, or in the area of community detection~\cite{Wu2004}.
A more detailed discussion, reviewing some of the notions of linear flows on networks, electrical quantities and classical relations to random walks can be found in Appendix~\ref{appendix:A}.
The links of resistor networks to random walks, commute times, and spectral properties of graphs have also been used for applications in data mining \cite{Saerens2004,Fouss2007}, and discussed in the context of convex optimization \cite{Ghosh2008} and graph sparsification \cite{Spielman2008}.
In all these contexts, however, the focus has still remained on the node space of the graph.
In contrast, here we are interested in defining \textit{relations between edges} in the network and using them for the analysis in the edge space of the graph.

The question of how edges influence each other arises naturally in electrical networks such as the power grid, in which it is important to assess the effect of an edge failure on the other edges in terms of the extra redistributed flow that those edges must carry.
This effect is quantified through the so-called line-outage distribution factor~\cite{Wood1996}.
We now present a graph-theoretical formulation of this concept and use it to construct an edge-to-edge matrix, the \textit{flow-redistribution matrix} that contains all such edge-to-edge dependencies.

A resistor network with weighted Laplacian $L$, given by~\eref{eq:laplacian}, and external current injection/ extraction $\mathbf{I}_{ext}$ is described by the network equations:
\begin{equation}\label{eq:Lap_equation}
  L \mathbf{V} = \mathbf{I}_{ext}.
\end{equation}
A set of node voltages $\mathbf{V}$ with zero mean and its corresponding edge currents $\mathbf{i}$ can be obtained by computing
\begin{eqnarray}
\mathbf{V} = L^\dagger \mathbf{I}_{ext}, \label{eq:pseudoinv_voltages}\\
\mathbf{i} =  GB^TL^\dagger \mathbf{I}_{ext}.\label{eq:pseudoinv_currents}
\end{eqnarray}
where $L^\dagger$ is the Moore-Penrose pseudoinverse of the Laplacian. For a detailed discussion see Appendix~\ref{appendix:A1}.

\begin{figure*}[tb]
  \centering
  \includegraphics[width=\textwidth]{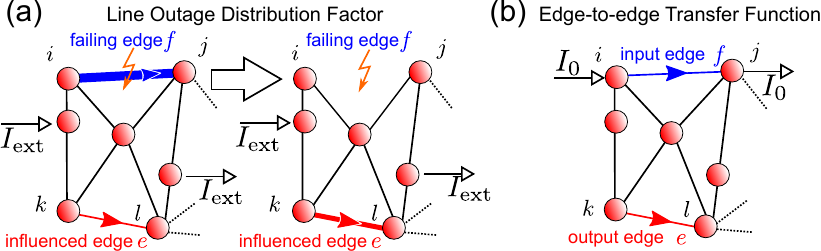}
  \caption{Schematic description of the line outage distribution factor (columns of the flow-redistribution matrix) and the edge-to-edge transfer function. (a) Line outage distribution factor: a line failure of edge $f$ will influence the flow on other edges in the network, as illustrated here for edge $e$. (b) Edge-to-edge transfer function: an ideal unit current injection along an edge $f$ induces flows in the network, as depicted here for edge $e$. }\label{fig:attached_current_source}
\end{figure*}

Consider now a line outage event: an edge $f$ fails and the flow redistributes through the network (see Figure \ref{fig:attached_current_source}a).
The redistributed flow can be calculated easily as follows.
The Laplacian matrix $ \widehat L_f$ of the new network after the failure of edge $f$ is:
\begin{equation}\label{eq:lap_matrix_failed_edge}
  \widehat L_f = L - g_f \mathbf{b}_f \mathbf{b}_f^T.
\end{equation}
Applying a generalised version of the Sherman-Morrison-Woodbury formula for the pseudoinverse~\cite{Meyer1973}, the new voltages are:
\begin{equation}
  \widehat{\mathbf{V}} = \widehat L_f^\dagger \mathbf{I}_{ext} = \left( L^\dagger + \frac{L^\dagger \mathbf{b}_f g_f \mathbf{b}_f^T L^\dagger}{1 - g_f \, \mathbf{b}_f^TL^\dagger \mathbf{b}_f}\right)  \mathbf{I}_{ext}.
\end{equation}
The change in the node potentials is then:
\begin{equation}
\label{eq:deltaV}
  \Delta_f \mathbf{V} = ( \widehat L_f^\dagger -  L^\dagger)\mathbf{I}_{ext} = \frac{L^\dagger \mathbf{b}_f g_f \mathbf{b}_f^T L^\dagger}{1 - g_f \, \mathbf{b}_f^TL^\dagger \mathbf{b}_f} \mathbf{I}_{ext}.
\end{equation}
Note that $\mathrm{i}_f$, the current on edge $f$ \textit{before} its failure, is:
\begin{equation}
\label{eq:i_edge}
 \mathrm{i}_f = g_fv_f = g_f \mathbf{b}_f^T \mathbf{V} = g_f \mathbf{b}_f^T L^\dagger \mathbf{I}_{ext}.
\end{equation}
Using~\eref{eq:pseudoinv_currents}~and~\eref{eq:deltaV}--\eref{eq:i_edge}, the
$E \times 1$
vector of changes in the edge currents when edge $f$ fails can be written as:
\begin{equation}
\label{eq:vector_LODF}
  \Delta_f \mathbf{i} =
\left [ \frac{GB^TL^\dagger \mathbf{b}_f}{1-g_f \, \mathbf{b}_f^TL^\dagger \mathbf{b}_f} \right ] \mathrm{i}_f \equiv  \mathbf{k}_f \;\mathrm{i}_f,
\end{equation}
In the electrical engineering literature, the vector $\mathbf{k}_f$ is called the \textit{line outage distribution factor} (LODF) for edge $f$.

Intuitively, the line outage distribution factor is a measure of the \textit{edge-to-edge} dependency in terms of the flow redistribution following an edge failure.  Crucially, $\mathbf{k}_f$  is \textit{independent} of the injected current pattern $ \mathbf{I}_{ext}$.
If we consider the effect of each of the $E$ edges failing in turn, we get the corresponding vectors $\mathbf{k}_i$, which we assemble into the \textit{flow-redistribution matrix}:
\begin{equation}
K_{E \times E} \equiv [\mathbf{k}_1 \cdots \mathbf{k}_E],
\end{equation}
which describes the edge-to-edge sensitivity under all possible single edge failures.
Again, the flow redistribution matrix is independent of the particular current injection and $K$ describes a \textit{topological property} of the system: the edge-to-edge influence under a perturbation of the flows on the links.

We remark that the $f$-th component of $\Delta_f\mathbf{i}$ in~\eref{eq:vector_LODF} (and hence the diagonal entries of $K$) does not correspond to the (trivial) change in current on the failed edge.
We will show below that these entries convey information which can be directly related to structural properties of the failing edge.

\subsection{Decomposing the flow redistribution matrix}

The matrix $K$ is one of the key ingredients for our edge-centric network analysis.
However, to gain a deeper understanding, it is insightful to pause here to discuss some important graph theoretical notions underlying the structure of $K$.

Note, that the flow redistribution matrix can be factorised as the product of two matrices with specific graph-theoretical meaning as follows.
Consider a network with weighted Laplacian $L$ and assume we inject and extract a current $I_0$ at the tail and head of edge $f$, i.e., $\mathbf{I}_{ext} = I_0 \mathbf{b}_f$ (see Figure \ref{fig:attached_current_source}b).
Equation~\eref{eq:pseudoinv_currents} shows that such an injection/extraction of current across edge $f$ induces the following current flows in the rest of the network:
\begin{equation}\label{eq:ETM_vector}
  \mathbf{i}_{[f]} = \left [GB^TL^\dagger \mathbf{b}_f \right ] I_0 \equiv I_0 \, \mathbf{m}_f.
\end{equation}
The edge vector $\mathbf{m}_f$ is a \textit{transfer function} relating the injection/extraction of the current $I_0$ at edge $f$ to the currents induced on all other edges.
The matrix compiling all transfer function vectors is the \textit{edge-to-edge transfer function matrix}:
\begin{equation}\label{eq:ETM}
M_{E \times E} \equiv [\mathbf{m}_1 \cdots \mathbf{m}_E] = GB^TL^\dagger B,
\end{equation}
where an entry $M_{ef}$ describes how an `input' unit current injected/extracted at (the endpoints of) edge $f$ is translated into an `output' current flowing at edge $e$.
Using~\eref{eq:vector_LODF}, we rewrite~\eref{eq:ETM_vector} in terms of the line outage distribution factor vector $\mathbf{k}_f$ as:
\begin{equation}
  \mathbf{i}_{[f]} =   I_0 \left [1-g_f \mathbf{b}_f^TL^\dagger \mathbf{b}_f \right ] \mathbf{k}_f   \equiv  I_0 \,  \varepsilon_f \, \mathbf{k}_f ,  \label{eq:def_ETF_embed}
 \end{equation}
where we have defined the \textit{edge embeddedness}, $\varepsilon_f$.

With these definitions, the flow-redistribution matrix can be rewritten as
\begin{equation}\label{eq:LODF}
  K  = M \, [\mathrm{diag}(\boldsymbol{\varepsilon})]^{-1},
\end{equation}
where $\boldsymbol{\varepsilon}$ is the vector of edge embeddednesses \footnote{The columns of the flow-redistribution matrix are undefined for edges with zero embeddedness. As will be become clear in Section~\ref{sec:Embeddedness}, if such an edge fails, the effect can be trivially understood by considering the related subgraphs independently.  Hence, we only consider examples in which the flow-redistribution matrix is well defined.}.
From this decomposition, it becomes clear that an edge failure will affect the edges in the graph in a similar way as if an additional source were attached to the failing edge with strength inversely proportional to the embeddedness of this edge.

The matrices $K$ and $M$ and the vector \E\,  constitute the main object of our work as graph-theoretical tools for the analysis of edge-to-edge relations, as shown below in detail.

\subsubsection{The edge-to-edge tranfer function matrix $M$}
As discussed above, the edge-to-edge tranfer function matrix $M$ describes the input-output relations in the edge space of the graph.
However, it has further important graph-theoretical properties of interest in different fields: it can be regarded as a discrete Green's function on the edge space of the graph, and it also appears in contexts such as graph sparsification~\cite{Spielman2008}.

Graph-theoretically, $M$ defines an orthogonal projection onto the \textit{weighted cut space} of the graph (see Appendix~\ref{app:B}). The weighted cut space, which is defined as the range of $GB^T$ or the column space of $K$~\eref{eq:LODF} (provided no edge has zero embeddedness), establishes the linear combinations of weighted edge vectors that disconnect the network.
Hence the action of $M$ has a purely graph-theoretical interpretation: it finds the projection of an `input' edge current (or combinations of those) onto the space of weighted cuts, thus evaluating how much of the input current gets distributed onto the weighted cuts disconnecting the network.

The matrix $M$ can also be understood in terms of effective resistances and commute times.
Consider edge $e$ linking nodes $i$ and $j$, and edge $f$ linking nodes $k$ and $l$.
From \eref{eq:ETM},
\begin{eqnarray}
 M_{ef} =& g_e (L^\dagger_{ik} - L^\dagger_{il} + L^\dagger_{jl} - L^\dagger_{jk}) \nonumber \\
 = & \frac{g_e}{2}(R_{jk}-R_{ik}+R_{il}-R_{jl})  \label{eq:ETM_eff_resistances}  \\
 =& \frac{\pi_e }{4}\left( (T_{jk}-T_{ik})+(T_{il}-T_{jl})\right)  \label{eq:commutetimes}
\end{eqnarray}
where $R_{ij}$ is the resistance distance and $T_{ij}$ is the commute time between two nodes $i,j$ (see Appendix \ref{appendix:A3}).
Thus $(T_{jk} - T_{ik})$ is the difference of commute times to nodes $i$ and $j$ when starting from node $k$, and $(T_{il} - T_{jl})$ is the difference of commute times to nodes $i$ and $j$ when starting from node $l$.
Here, $\pi_e = g_e/\mathrm{trace}(G)$ is just the probability of a random walker crossing edge $e$ (in any direction) at stationarity in the original network.
From this point of view, the edge-to-edge transfer function compares the difference in commute times to the two nodes of the `output' edge $e$ as observed from the two nodes of the `input' edge $f$.
A similar formula to~\eref{eq:ETM_eff_resistances} for the flow-redistribution matrix $K$ can also be given~\cite{Lehmann}.

The relationship between the flow-redistribution matrix $K$ and the edge-to-edge transfer function matrix $M$ is subtle.
While $M$ describes how a current injected/extracted at an edge translates into currents at all edges, the flow-redistribution matrix describes the relative dependency of edge flows under edge failure.
The edge-to-edge transfer function appears naturally as the flow-redistribution matrix of a partial \textit{$\alpha$-line failure}.
Assume that instead of a complete failure of edge $f$, its conductance is fractionally reduced by $\alpha g_f, \, \alpha \in [0,1]$. From~\eref{eq:lap_matrix_failed_edge}, the Laplacian after such an $\alpha$-line failure is $\widehat L_f (\alpha) = L - \alpha g_f \mathbf{b}_f \mathbf{b}_f^T$.
Assuming the same $\alpha$ applies to all edges, the flow-redistribution matrix for the $\alpha$-line failure is:
\begin{equation}
  K (\alpha)  = \alpha M \left [I - \alpha \,  \mathrm{ diag}(M_{ee}) \right]^{-1}.
\end{equation}
For small $\alpha$, this expression can be linearised to give:
\begin{equation}
 K(\alpha) \approx  K(0) + \left.  \frac{dK(\alpha)}{d\alpha}\right |_{\alpha = 0} \alpha = 0 + \left. M \left [I - \alpha\, \mathrm{ diag}(M_{ee}) \right]^{-2}\right .\Bigl|_{\alpha = 0} \alpha = \alpha M.
\end{equation}
Therefore $M$ is the slope with which small conductance fluctuations at each edge affect the flow on the other edges.

\subsubsection{The edge embeddedness $\varepsilon$}\label{sec:Embeddedness}

The embeddedness of edge $e$ that we defined in~\eref{eq:def_ETF_embed} can be rewritten as:
\begin{equation}
\label{eq:embed_def}
  \varepsilon_e = 1-g_e \, \mathbf{b}_e^TL^\dagger \mathbf{b}_e = 1 - M_{ee} = 1 - g_e R_e,
\end{equation}
where $M_{ee}$ is the corresponding diagonal element of $M$ and $R_e \equiv R_{h(e) t(e)}$  is the resistance distance \eref{eq:resist_dist} between the two endpoints of edge $e$.
Expression~\eref{eq:embed_def} makes again clear that the resistance distance along an edge, $R_e$, is not the same as its local, `physical' resistance, $r_e = 1/g_e$. In fact, the edge embeddedness measures how close $R_e$ and $r_e$ are.

It is well known from Rayleigh's Monotonicity law~\cite{Doyle1984} that $R_e \leq r_e$, with equality only if edge $e$ is part of no graph cycle, i.e., if $e$ accounts for the only path between $t(e)$ and $h(e)$. Indeed, $R_e$ can always be written as the local resistance $r_e$ in parallel to a resistance $R_\mathrm{rest}$ stemming from the rest of the network:
\begin{equation}
  \frac{1}{R_e} = \frac{1}{R_\mathrm{rest}}+\frac{1}{r_e}.
\end{equation}
Intuitively, $R_\mathrm{rest}$ will be small if the network has many alternative paths (i.e., cycles) with low resistance connecting $h(e)$ and $t(e)$.
Hence for $\varepsilon_e$ to be large, edge $e$ should participate in many cycles of short weighted length, i.e., it should be highly `embedded' and not crucial for the weighted cuts of the graph. On the other hand, a small $\varepsilon_e$  indicates that the edge participates in few cycles of small weight in the network. Such an edge would have a major influence on the induction of cuts in the network and is key in providing a connection that keeps the network connected. It is important to remark that the edge embeddedness is \textit{not} just another measure of betweenness centrality, as can be easily seen in a variety of examples discussed in Appendix~\ref{app:C}).

Some complementary interpretations of the embeddedness are also worth noting briefly. In terms of random walks, \eref{eq:commutetimes}~and~\eref{eq:embed_def} allow us to write the embeddedness of an edge $e$ with tail node $i$ and head node $j$ as:
\begin{equation}
  \varepsilon_e = 1 - \pi_e  \frac{T_{ij}}{2} = 1 - \frac{T_{ij}}{2\tau_e},
\end{equation}
where $\tau_e$ is the expected time for a random walker to return to edge $e$.
Thus the embeddedness compares the expected return time of a random walker to an edge and the commute time between the two edge endpoints.
Furthermore, for unweighted graphs: (i) the embeddedness of an edge is the probability that the edge is not found in a spanning tree selected randomly with uniform probability, which
follows directly from the interpretation of the resistance distance in terms of spanning trees~\cite{Doyle1984};
(ii)  the embeddedness of an edge provides a measure of how global the influence of a current injection along edge $e$ is, which follows from the fact that $M$ is symmetric and idempotent (see eq.~\eref{eq:idempotent}) and $M_{ee}$ is equal to the squared $L_2$ norm of the columns of $M$~\cite{Spielman2008}.

\section{Using edge-to-edge measures for network analysis}

Let us now use the flow-redistribution matrix $K$, and the edge embeddedness $\varepsilon$ defined above to provide an edge-centric analysis of networks.
To aid us in our network-theoretic analysis, we draw upon tools from community detection.
Specifically, we use the recent method of \textit{Markov stability of graph communities} \cite{Delvenne2010, Delvenne2012, Lambiotte2009} to find relevant groupings of edges, by interpreting the flow-redistribution matrix as the adjacency matrix of an effective edge-to-edge network. Thus we do not seek to partition the original graph into distinct node communities but rather aim at \textit{grouping edges according to their influence on each other}.
Note that the specific choice of community detection method is not essential, and any other community detection method can be used in conjunction with our edge-to-edge measures.
However, Markov stability is particularly useful for our purposes since it intrinsically scans across scales, thus enabling the detection of communities that include long-range or non clique-like structures, which can escape detection by other commonly used methods~\cite{Schaub2012,Schaub2012a}.
The relevant partitions are then selected based on their robustness properties. 
Within the framework of Markov stability, we consider partitions to be relevant only if they are robust to variability both to the optimization of the cost function and to the parametric dependence on the scale given by the Markov time, i.e., the robustness is assessed via the variation of information of the found solutions at each Markov time, as well as the persistence of a partition throughout Markov time (see \cite{Delmotte2011,Schaub2012}).

\subsection{A simple constructive example: a ring of small-worlds}
\begin{figure*}[tb!]
  \centering
  \includegraphics[width=\textwidth]{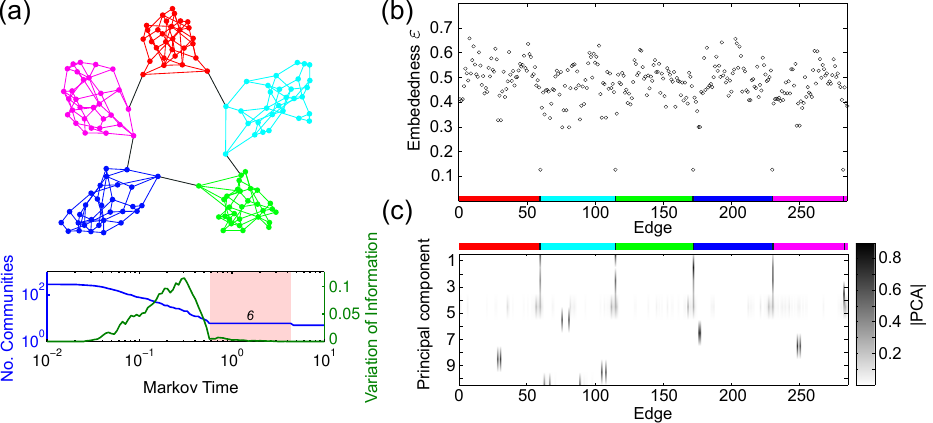}
  \caption{Edge-to-edge analysis of a ring of small-worlds. (a) The network analyzed with edges coloured according to the community structure found in the flow-redistribution matrix using the Markov stability method. The partition into 6 communities is stable over a long span of the Markov times with vanishing variation of information, thus signalling its robustness. (b) Embeddedness of the edges in the network. (c) Heat map of the first ten PCA components of the flow-redistribution matrix. Note that the edges linking the small-worlds: are grouped together in one community in (a); have low embeddedness in (b); and concentrate a large weight of the dominant principal components in (c).}
  \label{fig:2}
\end{figure*}

To illustrate our analysis, consider a network of $N=150$ nodes in which 5 small-world~\cite{Watts1998} subgrids of 30 nodes each are coupled in a ring-like structure (see Figure \ref{fig:2}a and Ref.~\cite{Schaub2012} for details).
Intuitively, the links between the individual subgrids are most critical for the flows traversing the system.
In case of failure, the inter-grid links will have an effect not only on the flow distribution inside the sub-grids but more importantly on the other inter-grid couplings, since all the flow that went through a particular inter-grid link would have to be `re-routed'.
Such a failure might thus lead to an overloading of another distant inter-grid link---a \textit{non-local} effect that does not follow trivially from the pattern of immediate node adjacencies.
In power grids, the significance of this event is obvious: an overloading of another line might in turn lead to another line failure possibly resulting in a rapid cascade of failures and a blackout of the system.
This intuitive picture can be captured quantitatively with our analysis, as shown in Figure \ref{fig:2}.
Figure \ref{fig:2}b shows that the links between the sub-grids show the smallest values
of embeddedness in the network, as expected.

In order to detect edge-to-edge influences, we analyse the community structure of graph edges using Markov stability~\cite{Delvenne2010, Delvenne2012, Lambiotte2009} on the weighted, directed adjacency matrix of absolute values of the flow-redistribution matrix (with removed diagonal).
We find a robust partition into six communities: 5 communities correspond to the subgrids, and \textit{all the links between subgrids are grouped into another community} (Figure \ref{fig:2}a).
As stated above, the robustness of the partition is to be understood here (and in the examples below) in two ways: (i) robustness with respect to the optimization of the cost function (Markov stability) of the partitioning at the particular Markov time (which is seen as a low value of the variation of information calculated from an ensemble of  runs of the Louvain algorithm); and (ii) robustness with respect to the parametric dependence on the Markov time, i.e., the partition is persistent in time as shown by the existence of a long plateau across Markov time (see Fig.~\ref{fig:2}a).

As one might expect, edges within a subgrid are clustered together, as their influence is mostly constrained to their local subgrid.
The fact that the inter-grid links form one community means that their influence on each other is very strong.
These edges also possess a relatively strong influence on the adjacent subgrids (as they can `disconnect' them) but their relative influence on each other is even stronger.
In fact, the magnitude of the line outage distribution factor between two of these edges is exactly one, indicating that in case of line failure the other inter-grid edges would be maximally affected.
The use of community detection in combination with the flow-redistribution matrix thus reveals \textit{non-local} properties of the network.
In the context of power grids, discovering such structural features could complement percolation-based node-centric analyses (see e.g. Brummit et. al \shortcite{Brummitt2012}) and provide input to load-flow based cascading failure models~\cite{Lehmann2010}.

The above community analysis is confirmed through a complementary principal component analysis (PCA) of the flow-redistribution matrix $K$ (Figure \ref{fig:2}c).
As discussed above, the range of $K$ (and hence its principal components) lie in the weighted cut space of the graph.
Therefore, PCA reveals the most important weighted cuts in the network with respect to flow redistribution.
Figure \ref{fig:2}c shows that the first principal components only have components involving the inter-subgrid couplings, confirming the results of our community detection.
In all the examples below, we have systematically carried out this PCA analysis (not shown), which similarly confirm the results obtained with the edge embeddedness and Markov Stability community detection.

\section{Applications to real-world networks}
We now consider several real-world examples to illustrate the general applicability of our edge-centric tools.
Our aim here is not to perform an in-depth analysis of each of these systems, which would be beyond the scope of this paper, but rather to highlight different aspects of the edge-to-edge measures introduced above.

\subsection{The Iberian Power Grid}
\begin{figure*}[tb!]
  \centering
  \includegraphics[width=\textwidth]{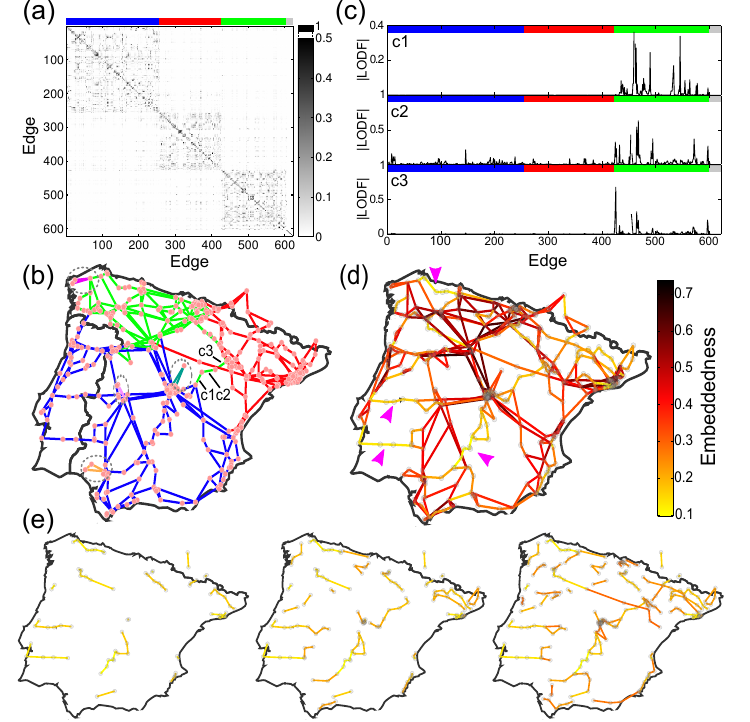}
  \caption{Analysis of the Iberian Power Grid.  (a) flow-redistribution matrix ordered according to the community structure found with the Markov stability method.  (b) Map of the Iberian Power Grid with colours denoting edge communities. The community structure displays \textit{non-local} structure: the edges c1--c3 are grouped with the north-west (green) community, although these edges lie between the north-east (red) and central-south (blue) communities and have no direct connection with the north-west (green) community. Small local circles (encircled with gray dotted lines), form their own isolated communities, i.e., they are effectively 'decoupled' from the rest of the network. (c) Influence of edges c1--c3 on all other edges in the network as measured by the magnitude of the line outage distribution factor (LODF). (d) Edge embeddedness of all the edges in the network. There are several weakly embedded paths of lines (marked with magenta arrows), e.g., those connecting the center and south of Portugal
with Spain; the lines going from the centre to the south, from Madrid towards Andalusia; or the lines connecting Asturias and Galicia along the North-Northwest coast (see \url{http://en.wikipedia.org/wiki/List_of_power_stations_in_Spain}). (e) Weakly embedded edges in the Iberian Power Grid. From left to right, the lowest $10\%, 20\%$ and $35\%$ embedded edges and associated nodes.}
  \label{fig:3}
\end{figure*}

Our first example is the Iberian subnet of the European Power Grid~\cite{Rosas-Casals2007,Sole2008,Schaub2012}, which consists of 403 nodes corresponding to generators and substations and 622 edges representing high-voltage transmission lines. Our description of power systems as resistor networks corresponds to the so-called DC power flow approximation, a common linearised representation of the non-linear load-flow equations around a reference state.
Beyond ascertaining the $N-1$ robustness against failure propagation~\cite{Wood1996}, we apply here our network-theoretic analysis to reveal (non local) edge-to-edge features in this network. 

Our community detection analysis finds a robust partition that splits the edges into three main communities, as shown in Figure \ref{fig:3}a-b.
Interestingly, this partition uncovers non-local relationships between the edges: the transmission lines that connect the north-east with the central part of the grid (edges c1-c3 in Figure \ref{fig:3}b), roughly going from Saragossa towards Madrid, appear to be strongly linked to the north-western part of the grid and form part of this community (green).
Figure \ref{fig:3}c confirms this finding:  the influence of edges c1-c3 are much more significant on the north-west (green) community.
This behaviour follows from the fact that edges c1-c3 are part of a long loop going from the northwest eastwards; connecting to the center via a southern branch containing edges c1-c3; and eventually going back to the northwest.

An analysis of the embeddedness of the edges in the Iberian grid is shown in Figure~\ref{fig:3}d-e.
As we might expect from our previous analysis, edges c1-c3 are only weakly embedded in the graph. 
Note also that the lines connecting the center and south of Portugal with Spain show very small embeddedness due to the lack of alternative routes.
A similar observation applies to the line leading from Madrid towards the south and the line connecting Asturias and Galicia in the Northwest coast.
All of these lines are indicated with magenta arrows in Figure~\ref{fig:3}d.
Interestingly, several of these lines are associated with relatively new solar plants (see \url{http://en.wikipedia.org/wiki/List_of_power_stations_in_Spain}).
An additional assessment of the importance of individual lines is shown in Fig.~\ref{fig:3}e, in which the skeleton of increasingly embedded lines of the Iberian grid is displayed.

\subsection{Traffic networks}
As a second example, we consider traffic networks corresponding to parts of the street networks of London, Boston and New York~\cite{Youn2008}.
We analyze the networks reported in Ref.~\cite{Youn2008} (data kindly provided by H. Youn), in which the nodes correspond to street intersections and the edges are principal roads as classified by Google Maps.
In our analysis we assume the streets to be undirected and the edge weights correspond to the number of street lanes.
In these systems, currents can be naturally identified with traffic flows and voltages with delays, although the relationship between flows and delays is in general non linear (see \cite{Youn2008} and references therein).
Hence, our analogy with a linear resistor network amounts to assuming a socially optimal behaviour for all drivers, and in particular ``Braess paradox'' \cite{Youn2008,Witthaut2012} cannot arise in our context.
However, based on our simplified linear model, we use the flow-redistribution matrix and related measures to perform a coarser, topological analysis of traffic flows independent of patterns of injected flow.
We can thus assess the relative interdependence and importance of the edges (roads) with respect to any (linear) traffic flow, rather than focussing on the influence of an edge for a particular source-target pair.

\begin{figure*}[tb!]
  \centering
  \includegraphics[width=\textwidth]{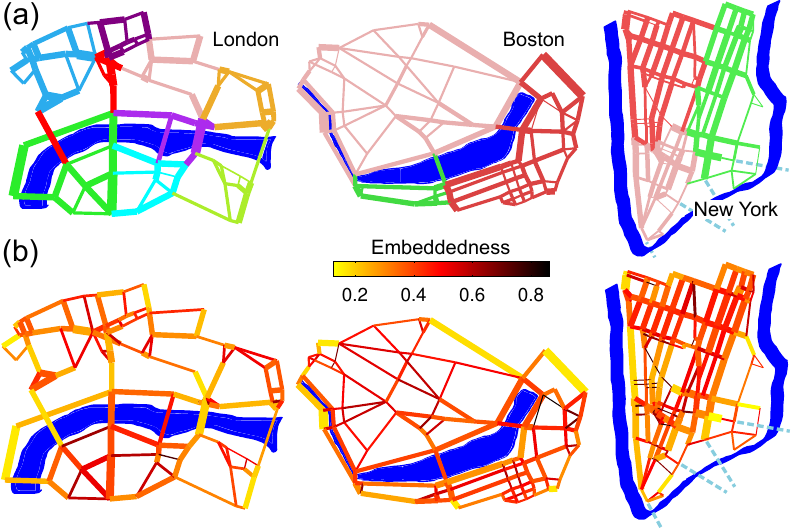}
  \caption{Analysis of Urban Street Networks of London (82 nodes, 130 edges), Boston (88 nodes, 155 edges), and New York (125 Nodes; 217 edges).  Nodes correspond to intersections and edges to (undirected) streets weighted according to the number of lanes. Light blue dashed lines indicate some connecting roads not part of the analyzed network. (a) Communities of streets (denoted by different colours) found from the analysis of the flow-redistribution matrix with the Markov stability method. The streets within each community have a strong influence on each other. Unlike Boston and New York, we detect non-local community structure in the streets of London (red community). (b) Embeddedness of the edges in the street networks. The mean embeddedness in London, $\langle \varepsilon \rangle_\mathrm{London} =0.377$, is lower than for the US cities ($\langle \varepsilon \rangle_\mathrm{Boston} =0.439, \langle \varepsilon \rangle_\mathrm{New York} =0.429$), mainly due to the more grid-like structure of the principal
roads in the US
street networks. Note the low embeddedness of most bridges (or continuation streets) in London and the existence of a core of highly embedded streets at the centre of Lower Manhattan.}
  \label{fig:4}
\end{figure*}

Figure~\ref{fig:4} displays the results of our community detection algorithm on these street networks based on the edge-to-edge flow-redistribution matrix.
In the case of London, we find a robust partition into nine communities of streets, eight of which correspond to well delimited city areas while the ninth is a non-local community of edges comprising two alternative main north-south routes across the Thames: Waterloo Bridge and Farringdon Street, which is a continuation of Blackfriars Bridge.
Our analysis indicates that these two routes are therefore strongly coupled in terms of flow redistribution.
For the two American cities, such non-local community structure is not observed, as could be expected given the more regular, grid-like structure of both networks.
In the case of New York, we obtain a robust partition into three communities of streets corresponding approximately to Lower Manhattan/Financial district in the south; Kips Bay/Lower East Side/East Village on the East side; and Greenwich Village/Chelsea on the West side.
Similarly, Boston is split into three communities of streets corresponding to Back Bay/Downtown/Beacon Hill; a second community extending over Cambridge; and a third, smaller community comprising the Boston University area and Harvard Bridge over the Charles.

The study of the edge embeddedness reveals further differences between the cities.
In particular, London and New York present the most dissimilar profiles of $\boldsymbol{\varepsilon}$: London has the lowest mean embeddedness with a significant tail of streets with low $\varepsilon$, while New York has the broadest distribution of $\varepsilon$.
The edge embeddedness in New York markedly increases as we go towards Chinatown/Little Italy/Canal Street, where we find a central core of highly embedded streets.
This is expected from the grid-like structure of the street network one typically encounters in American cities, which by construction provides many alternative paths to most locations in the network.
New York also has a set of streets with low embeddedness mostly in the periphery.
The presence of low $\varepsilon$ edges at the boundaries of the graph is expected since the flows at the boundaries have fewer alternative paths to be redistributed.
Studying the relevance of such low peripheral $\varepsilon$ on larger street networks that have not been artificially `cropped' will be the subject of future work.
Interestingly, the presence of `internal boundaries' can also induce low edge embeddedness.
An example for such a street with low $\varepsilon$ is the Lincoln Highway/West Street on West Lower Manhattan, which has the Hudson River as a natural boundary.
In the case of London, a significant fraction of the streets with low $\varepsilon$ lies in the north-south direction, connecting the areas south of the river Thames with the northern part of the network.
Most of these roads correspond to bridges, which are bottlenecks in the real street network.
In fact, all but one bridge have $\varepsilon$ below the mean, including Waterloo Bridge, London Bridge and Westminster Bridge with particularly low scores.
The street network of Boston shows a less extreme grid-like structure than that of New York and falls therefore somewhere in between London and New York (see Figure \ref{fig:4}).

\subsection{Neuronal network of C. Elegans}

\begin{figure*}[tb!]
  \centering
   \includegraphics[width=\textwidth]{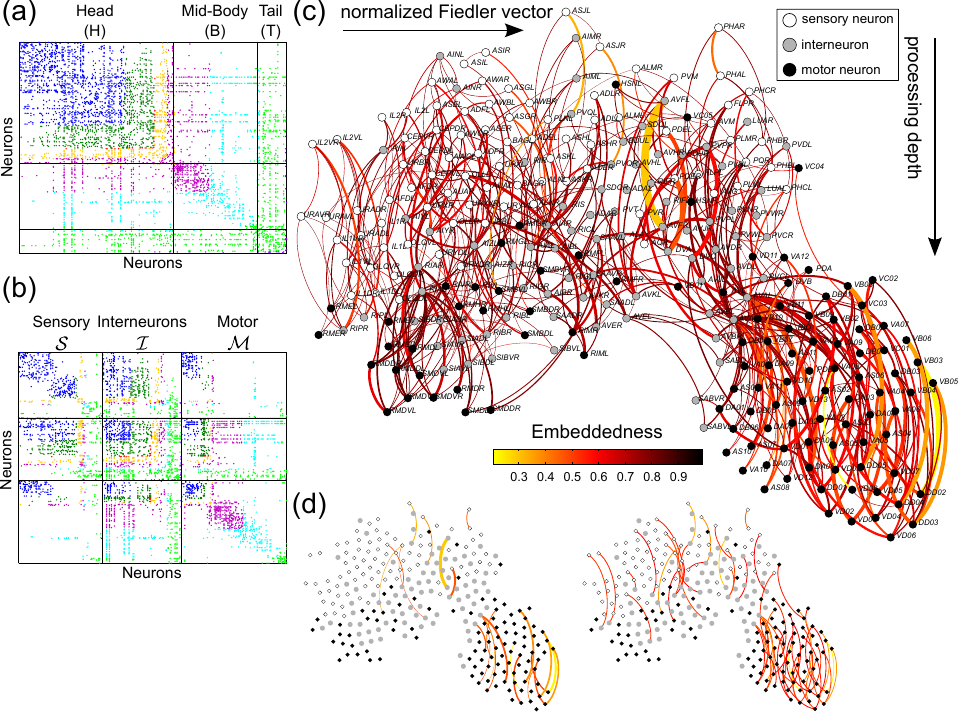}
  \caption{Analysis of the neural network of  \textit{C. elegans}. The edges of this network (synapses and gap junctions) were found to belong to 8 robust communities (denoted by different colours in (a) and (b)) according to the analysis of the flow-redistribution matrix using the Markov stability method. (a)  Visualisation of the edge communities in the adjacency matrix ordered according to body position (anteroposterior order). (b) Visualisation of the edge communities in the adjacency matrix ordered according to functional categories: sensory neurons \mS, interneurons \mI\, and motor neurons \mM. An anteroposterior ordering is applied within each group. (c) Embeddedness of the edges in the \textit{C. elegans} neural network. Neurons are coloured according to type: sensory neurons  (white), interneurons (gray), and motor neurons (black). For the visualisation of the network we used the planar display suggested by Varshney et. al \shortcite{Varshney2011}: vertical axis corresponds to the position of the
neuron in the signaling pathway (sensory neurons tend to be at the top, motor neurons at the bottom); horizontal axis is the normalised Fiedler vector (which tends to group nodes with more connections to each other closer in space). In this visualisation, we see that the embeddedness grows as the processing depth increases: synapses between sensory neurons (upstream) tend to be more embedded, while edges linked to motor neurons (downstream) tend to be less embedded. (d) This observation is also confirmed by the skeleton of weakly embedded edges in the neuronal network of \textit{C. elegans}: the connections with the lowest 1 percent (left) and 3 percent (right) edge embeddedness.}
  \label{fig:5}
\end{figure*}

Our final example is the neural network of the worm \textit{C. elegans}, one of the few model organisms for which the entire neural wiring is almost completely available.
Here we use the strongly connected giant component of the network of gap junctions and chemical synapses~(recently enlarged and curated by \cite{Varshney2011}, \url{http://web.mit.edu/lrv/www/elegans/}), which consists of 274 nodes (neurons) and 2253 edges (synapses and gap junctions), which we assume to be undirected.
An in-depth analysis of the functional and structural features of this neuronal network is beyond the scope of this paper --- for pointers to the vast and comprehensive literature on the subject, see, e.g.,~\cite{White1986, Varshney2011,Sohn2011} and references therein.

To display and interpret our results, we use the classification of neurons into body compartments and functional types in \url{http://www.wormatlas.org/neuronalwiring.html} \cite{Varshney2011}.
Position-wise, edges are denoted according to the compartment (head: H, mid-body: B, or tail: T) in which its end points lie, e.g., an HB edge connects the head and mid-body regions.
Type-wise, edges are denoted according to the type of neuron (sensory (\mS), interneuron (\mI) and motor (\mM)) that they connect, e.g., a \mS-\mI\, edge connects a sensory neuron to a motor neuron.

Figure~\ref{fig:5}a-b shows the eight communities of edges of this neuronal network, as obtained by analysing the flow-redistribution matrix with Markov stability.  Figure~\ref{fig:5}a shows the communities of synapses ordered according to body positions. As expected, the edge communities are closely linked to the body structure of the worm. More precisely, the communities are mainly centered around either head, mid-body, or tail positions, i.e. the core of each community comprises a group of either HH, BB, or TT edges.
Interestingly, the edges linking different regions tend to belong to communities centered around the region closest to the tail, e.g., HB edges tend to belong to body-centered communities, while HT edges belong to tail-centered communities (Figure~\ref{fig:5}a).
This indicates a `downstream' organisation in the way that synaptic changes affect other neurons: a synaptic failure will tend to cascade `downstream' from the head region, where most sensory neurons lie, towards the body and tail regions, where most interneurons and motor neurons lie.
In this sense, changes in sensory synapses 'upstream' tend not to affect other similar sensory synapses, and only affect synapses downstream.

Figure~\ref{fig:5}b shows the edge communities displayed in accordance with their associated neuronal types (\mS, \mI, \mM).
We find that the two communities of edges connecting to mid-body positioned neurons (magenta and cyan colours) correspond mainly to \mM-\mM\, or \mI-\mM\, edges. Hence these communities might be thought of as `downstream' executive communities.
On the other hand, the tail-centered community (light green) and one of the head communities (dark green) comprise mostly couplings from interneurons (of all types \mS-\mI, \mI-\mI, \mI-\mM), suggesting a key role of these edges, in agreement with the commonly accepted role of interneurons as controlling units in the neural circuitry.
The edge community (blue) with the strongest impact on the sensory modalities includes connections to all neuron types. In particular, the interneurons linked by the \mI-\mI\, edges in this blue community appear to have a central position in the network: they link from/to any edge community and neuron type, including a large number of connections to motor neurons. One may thus hypothesise that this group of interneurons interconnected by the \mI-\mI\ edges in the blue community acts as a control hub processing the inputs from sensory neurons and relaying it to motor neurons.

The edge embeddedness of the connections in the neuronal network of \textit{C. elegans} is shown in Figure \ref{fig:5}c-d. We find that the edge embeddedness decreases as the processing depth increases, i.e., edges with low embeddedness are predominantly located downstream, in the late stages of the processing hierarchy and connected to motor neurons (see Figures \ref{fig:5}d). This can be explained by the fact that motor neurons are essentially terminal nodes activated from upstream processing via only a few connections and, in this sense, they belong to weakly embedded `pathways'.
On the other hand, further up in the signaling chain (in synapses related to sensory neurons), very few edges have low embeddedness (Figure \ref{fig:5}d) indicating that signalling synapses are embedded in `circuits' with more alternative paths.
One notable exception is the connection between the interneurons AVFL and AVFR, which shows low embeddedness even if it is high up in terms of processing depth.
This low embeddedness reflects a lack of alternative paths for flow redistribution if this synapse fails.
Interestingly, the AVFL and AVFR neurons are thought to be involved as decision-making interneurons in the temporal coordination of egg-laying and locomotion of the nematode~\cite{Hardaker2001}.

\section{Discussion}
Analytical tools used to investigate complex networks have commonly adopted a node-centric perspective, aiming at the characterisation of individual nodes or of groups of nodes and their relations to each other.
In this paper, we have presented tools to characterise edge-to-edge relations inspired by the redistribution of flow induced by line failures.
We have shown that the flow-redistribution matrix is a topological descriptor of the network that can be used to quantify edge-to-edge relations induced by the flow redistribution after a single line failure. Further extensions of this work are currently under way to consider multiple line outages and the connection with cascading processes~\cite{Gueler2007}.

We have illustrated how flow-redistribution matrix can be decomposed into an edge-to-edge transfer function matrix, which describes how much the injection of flow at an edge translates in changes of flow in other edges, and a vector of edge embeddednesses, which describes how costly it is to transit between the two endpoints of each edge through alternative paths in the network.
Our analysis provides us with explicit network-theoretic interpretations of these edge-to-edge measures.
Adopting such an edge-based perspective can provide a complementary view of network properties and allows for a natural detection of structural features which may not be readily found by node-centric methods.

Importantly, the flow-redistribution matrix and the associated edge-to-edge transfer function matrix and embedddedness vector \E\, take into account non-local properties of the graph and go beyond local adjacency relations between edges, as represented by the line graph~\cite{Evans2009, Ahn2010}.
This fundamentally non-local nature of our measures emanates from the fact that their graph theoretical description is underpinned by the pseudoinverse of the Laplacian.
The pseudoinverse of the Laplacian incorporates global properties of the graph and serves to link our measures to other (graph) theoretically relevant properties such as the resistance distance, commute and hitting times of random walks as well as graph embeddings. As discussed in Appendix~\ref{appendix:A2}, there exist efficient algorithms for the computation of these measures which are equivalent to the solution of a linear sparse system.

The examples presented above highlight how our edge-based measures are able to detect relevant structural features with an impact on the dynamics of the respective systems.
In addition, there are other applications in which adopting an edge-based perspective would appear natural, including metabolic control analysis, the structural analysis of biomolecules under bond fluctuations~\cite{Delmotte2011}, or financial networks, in which the disturbance of financial flows between different actors may have significant effects on different parts of the network.

\section*{Appendix}
\appendix
\section{Linear flows, electrical networks and random walks}
\label{appendix:A}

A large class of network processes can be modeled by linear dynamics on a network, described by state variables on the nodes and edges of a graph (c.f. the book by Strang \cite{Strang1986} for an insightful discussion and the reformulation of diverse problems in these terms).
Systems of this type include widely used models of spring-mass-damper networks of mechanical systems, as well as electrical networks and reversible Markov chains (i.e., random walks or diffusion processes on undirected networks), among many others.
In all cases, a constitutive relation links the flow along an edge with the node variables at its tail and head.
The simplest such relation is an Ohm-type law that establishes a linear relationship between the flow on the edge and the difference between the associated node variables.

\subsection{Linear flows on networks and electrical quantities}
\label{appendix:A1}
The canonical example of linear flows on edges is the electrical resistor network (and its analogy to random walks).
Henceforth, node variables are denoted by capital letters, while small letters are reserved for edge quantities.
In a resistor network,  the flows on the edges correspond to electrical currents driven by potential differences across the edges (Ohm's law). Each node $k$ in the network has an associated potential $V_k$ and the potential difference over edge $e$ is $v_e =V_{t(e)}- V_{h(e)}$.
Given the vector of node potentials $\mathbf{V}$, the vector of voltages across the edges is: $\mathbf{v} = B^T\mathbf{V}$.
The current on each edge is equal to the edge voltage times the conductance:
\begin{equation} \label{eq:Ohm}
\mathbf{i} = GB^T \mathbf{V} \quad \mathrm{(Ohm`s\: law)}.
\end{equation}
Furthermore, by Kirchhoff's current law (KCL), the in- and out-flow of currents at each node is balanced:
\begin{equation} \label{eq:KCL}
B \mathbf{i} = \mathbf{I}_{ext} \quad \mathrm{(Kirchhoff`s\:current\:law)},
\end{equation}
where $\mathbf{I}_{ext}$ is the vector of external currents injected into the nodes.~\footnote{If external voltage sources $\mathbf{v}_{ext}$ along the edges are present,  then $\mathbf{i} = G(B^T\mathbf{V} - \mathbf{v}_{ext})$.
We do not need to consider external voltage sources separately since each external voltage source can be transformed into its equivalent current source (Norton equivalent)}

The properties of the incidence matrix $B$ are directly connected with certain physical constraints.
First, the vector of ones $\mathbf{1}_{N \times1}$ is in the nullspace of $B^T$, consistent with KCL.
Hence $\mathbf{1}^T \mathbf{I}_{ext} = 0$ and the net injected current into the system must be zero.
Second, the nullspace of $B$ is the cycle space~\cite{Guattery1998,Godsil2001}, i.e., the space spanned by all cycle vectors.
Any (oriented) cycle in the graph can be represented by a vector $\mathbf{c}_{E \times 1}$ as follows: moving along the edges in the cycle, $c_e = 1$ if the edge direction is aligned with the direction of the cycle and $c_e = -1$ if it is opposite, with all other entries of $\mathbf{c}$ zero.
Then for any cycle, $\mathbf{c}^T \mathbf{v}= \mathbf{c}^TB^T\mathbf{V} = 0$, i.e., the voltage drop around any cycle in the graph must be zero \footnote{If magnetic fields need to be included in this formulation, they would be represented by additional current/voltage sources.}.
This is of course Kirchhoff's voltage law (KVL).

Combining \ref{eq:Ohm} and \ref{eq:KCL}, we get the well-known network equations relating input currents and node voltages:
\begin{equation}\label{eq:Lap_equation2}
  L \mathbf{V} = \mathbf{I}_{ext}.
\end{equation}
Using standard nodal analysis~\cite{Strang1986},  we must first solve for the potential of the nodes $\mathbf{V}$ in~\eref{eq:Lap_equation2} and then obtain the edge currents from~\eref{eq:Ohm}.
Equation~\eref{eq:Lap_equation} can always be solved, though not uniquely since $L$ is singular.
This corresponds to the fact that the node potentials have an arbitrary reference.
To fix a reference, the network is commonly \textit{grounded}, i.e., the potential of one (arbitrary) node is set to zero.
This leads to the definition of a $(N-1)$-dimensional \textit{grounded Laplacian matrix} obtained by deleting a row and the corresponding column~\cite{Yuan2013,Jadbabaie2004}.

Alternatively, a unique $\mathbf{V}$ can be obtained from \eref{eq:Lap_equation2} through the Moore-Penrose pseudoinverse of the Laplacian,  $L^\dagger$, which can be written as~\cite{Ghosh2008}:
\begin{equation}
  L^\dagger = \left(L+\frac{1}{N}\mathbf{1}\mathbf{1}^T\right)^{-1}-\frac{1}{N}\mathbf{1}\mathbf{1}^T.
\end{equation}
The particular vector of node potentials (and the corresponding edge currents):
\begin{eqnarray}
\mathbf{V}& = L^\dagger \mathbf{I}_{ext} \label{eq:pseudoinv_voltages2} \\
\mathbf{i} &= GB^TL^\dagger \mathbf{I}_{ext}.\label{eq:pseudoinv_currents2}
\end{eqnarray}
is the solution of~\eref{eq:Lap_equation2} with minimal $L_2$ norm, and $\mathbf{V}^T\mathbf{1} =0$. Hence the node potentials obtained have zero mean, i.e., the voltages are referred to the average potential~\cite{Jadbabaie2004}.

\subsection{Effective resistances and random walk interpretations}
\label{appendix:A3}
An important property of electrical networks is the effective resistance $R_{ij}$ between two nodes $i$ and $j$.
Physically, $R_{ij}$ is the potential drop measured when a unit current is injected at node $i$ and extracted at node $j$.  The effective resistance can be compactly written in terms of the Laplacian pseudoinverse~\cite{Ghosh2008}:
\begin{equation}
\label{eq:resist_dist}
 R_{ij} = (\mathbf{U}_i -\mathbf{U}_j)^TL^\dagger (\mathbf{U}_i -\mathbf{U}_j),
\end{equation}
where $\mathbf{U}_i$ is the $i$-th unit vector, with a one at the $i$-th coordinate and zeros in all other coordinates.
Clearly, $R_{ij}=R_{ji}$.
The effective resistance defines a distance metric on the graph~\cite{Klein1993} and is also commonly known as the \textit{resistance distance} (between two nodes).
For a detailed overview and additional interpretations of this quantity see~\cite{Ghosh2008} and references therein.
Note that $R_{ij}$ has a global dependence on the network as it takes into account all possible paths between $i$ and $j$. Therefore, even if nodes $i$ and $j$ are directly connected by an edge with conductance $g_e$, the effective resistance $R_{ij}$ will \textit{not} in general be equal to $1/g_e$.  This effect, induced by the presence of the network , underpins the concepts developed in this paper.

A broader, alternative perspective on the electrical formalism discussed above is provided by the theory of harmonic functions on a graph, which establishes a fundamental relationship between electrical networks and reversible random walks on a graph.
Detailed accounts of this topic are given in the books of Doyle and Snell~\cite{Doyle1984} and Aldous and Fill~\cite{Aldous2012}, amongst others.
In the context of random walks, the resistance distance is shown to be proportional to $T_{ij}$, the commute time of a random walker between nodes $i$ and $j$~\cite{Aldous2012,Lovasz1994,Ghosh2008}:
\begin{equation}
\label{eq:commute_time}
 R_{ij}= \frac{T_{ij}}{2 \, \mathrm{trace}(G)},
 \end{equation}
where $T_{ij}$ is the expected time for a random walker to return to node $i$ for the first time after starting from node $i$ and passing through node $j$.

The random walk picture also provides interpretations for the currents and voltages~\cite{Doyle1984}.
Let a unit current be injected into node $i$ and extracted at node $j$.
Then the current $\mathrm{i}_e$ corresponds to the net expected number of times a random walker which starts at node $i$ and walks until she reaches $j$ will cross edge $e$ in the defined orientation.
On the other hand, voltages can be interpreted as relative hitting probabilities.
Let a unit voltage be applied between nodes $i$ and $j$. Then the potential at node $k$ corresponds to the probability that a random walker starting from $k$ will hit node $i$ first before reaching $j$.

\section{Computational aspects of edge based measures}
\label{appendix:A2}
The computational cost of our method is dominated by the computation of the pseudoinverse of the Laplacian matrix, for which there are efficient methods~\cite{Bozzo2012}.
In fact, we do not need to compute the pseudoinverse explicitly, but rather solve a linear system of the form $Lx = b$.
As this system is usually sparse for many graphs, there exist very fast standard techniques to obtain the matrices $K$, $M$ and the vector of embeddedness \E.
In addition, there also exist fast algorithms to obtain approximately all currents and voltages in the network based on local averaging. The running time of such methods is $O(N+E)$ to obtain all voltages in the network~\cite{Wu2004}.
Hence all of our measures are computable by simple (sparse) matrix multiplications.
Alternatively, Spielman et. al have recently presented an efficient algorithm that allows the (approximate) computation of any resistance distance between any two nodes in the graph in $O(log(N))$ time \cite{Spielman2008}. Using this method in combination with formulas \eref{eq:embed_def} and \eref{eq:ETM_eff_resistances} can also facilitate the edge-centric analysis of very large networks in terms of the flow-redistribution.

\section{Additional Properties of the Edge-to-Edge Transfer Function matrix}
\label{app:B}

In the following, we elaborate on further properties and interpretations of the edge-to-edge transfer function matrix (see also Spielman \& Srivastava \shortcite{Spielman2008}).
First, $M$ is a projection (idempotent) matrix: $M^2=M$. To see this:
\begin{equation}
 M^2 = GB^TL^\dagger BGB^TL^\dagger B
		=GB^TL^\dagger L L^\dagger B = GB^TL^\dagger B = M,
\label{eq:idempotent}
\end{equation}
which follows from the definition of the pseudoinverse.
Second, all the eigenvalues of M are either zero or one.
To  prove this, consider the symmetrised matrix $\widetilde M = G^{-1/2}MG^{1/2} = G^{1/2}B^TL^\dagger BG^{1/2}$ and use the singular value decomposition of $B$. It is then easy to show that the eigenvalues of $M$ are $(N-1)$ ones and $(E-N+1)$ zeros (see Spielman \& Srivastava \shortcite{Spielman2008} for a different proof of the same results).

We can give a physical interpretation to these results as follows.
Since the graph has $N$ nodes and $E$ edges, we know there are $E-(N-1)$ independent cycles spanning the cycle space~\cite{Godsil2001,Guattery1998}.
Input currents that fall into the cycle space will balance and yield zero output, thus leading to the $E-(N-1)$ zero eigenvalues. Only inputs that lie in the orthogonal complement of the cycle space, the so called cut space \cite{Godsil2001,Guattery1998}, will yield a non-zero current output.
Let us call the current input orthogonal to the cycle space the effective input.
Conservation of flow implies that the effective input can only be redistributed in the network, i.e. the flow across any weighted cut can at most match this input.
In particular, the sum of the flows across any set of (weighted) cut vectors forming a basis for the weighted cut space has to be equal to the effective input.
This corresponds to the fact that the remaining $N-1$ eigenvectors of $M$ have unit eigenvalues.

\section{Additional Properties of Edge embedddedness and comparison to other centrality measures}
\label{app:C}

\begin{figure*}[tb!]
  \centering
  \includegraphics[width=\textwidth]{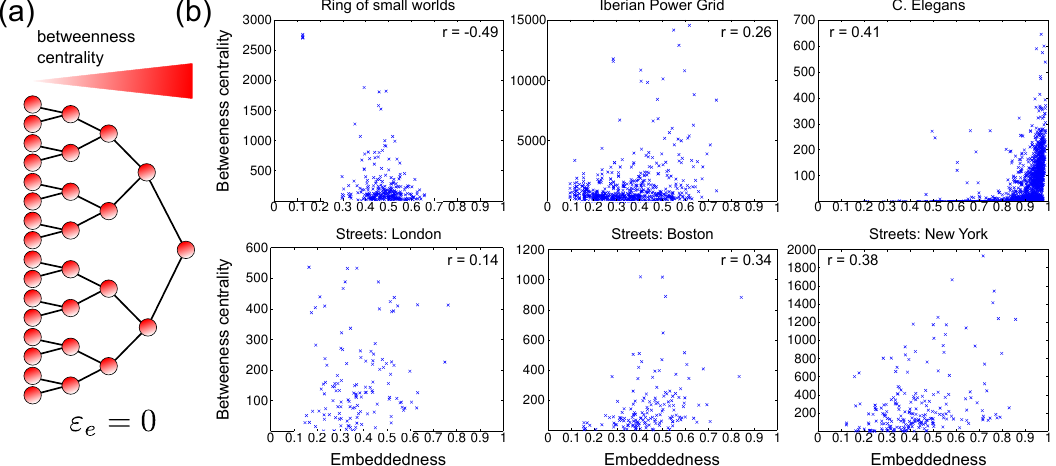}
  \caption{Comparison of the edge embedddedness and betweenness centrality. (a) Illustration of the difference between embeddedness and betweenness centrality with an unweighted hierarchical tree (the same argument applies to other centrality measures): while the betweenness of each edge depends on its position within the tree, the embedddedness is zero for all edges, independently of their position, since any edge failure will disconnect the graph. (b) Scatter plots of the embedddedness  against betweenness centrality for all edges of all the networks used in this work. Pearson correlation coefficients are displayed as r. No dependence between edge embeddedness and betweenness centrality is apparent.}
  \label{fig:6}
\end{figure*}

As discussed above, the embedddedness of an edge can be interpreted as measuring how much an edge forms a ``bottleneck'' in the network. Such a notion is also inherent in many (edge) centrality measures, which try to assess the importance of a particular node/edge in a network \cite{Freeman1978,Bonacich1987,Borgatti2005,Delvenne2011}.
The most prominent notion of edge centrality is arguably \textit{betweenness centrality} \cite{Freeman1977}, which measures how many times an edge participates in the shortest (geodesic) paths between any two nodes. Another popular centrality measure is based on random walks~\cite{Newman2005}.

It is important to note that the embeddedness of an edge presents significant differences with such measures of edge centrality.
While edge centrality measures assess how important a particular edge is for traversing between \textit{any} two nodes, the embeddedness measures how important an edge is for traversing between its two endpoints through alternative paths. Hence embeddedness incorporates the importance of cycles in the graph.
To illustrate this difference, consider a binary tree, as shown in Fig.~\ref{fig:6}a.
The closer we get to the root of the tree the higher the betweenness centrality of the edges will be.
In contrast, the embedddedness will be zero for all edges independent of their relative position, since the outage of any edge will disconnect the graph. Similar differences apply to random walk-based betweenness centrality~\cite{Newman2005}.

To give a more quantitative assessment of these differences, we display in  Figure~\ref{fig:6}b a numerical comparison between the betweenness centrality and the embeddedness of all edges for all the examples used in this work. No dependence between them is apparent, emphasizing that the embedddedness is an distinctive measure, different to betweenness centrality.

\bibliographystyle{nws}
\bibliography{references}

\begin{thebibliography}{}

\bibitem[\protect\citename{Ahn {\em et~al.}\relax, }2010]{Ahn2010}
Ahn, Yong-Yeol, Bagrow, James~P., \& Lehmann, Sune. (2010).
\newblock {L}ink communities reveal multiscale complexity in networks.
\newblock {\em Nature}, {\bf 466}(7307), 761--764.

\bibitem[\protect\citename{Albert \& Barab\'asi, }2002]{Albert2002}
Albert, R\'eka, \& Barab\'asi, Albert-L\'aszl\'o. (2002).
\newblock {S}tatistical mechanics of complex networks.
\newblock {\em Rev. mod. phys.}, {\bf 74}(Jan), 47--97.

\bibitem[\protect\citename{Aldous \& Fill, }2012]{Aldous2012}
Aldous, D., \& Fill, J. (2012).
\newblock {\em {R}eversible {M}arkov {C}hains and {R}andom {W}alks on
  {G}raphs}.
\newblock Book in preparation.
\newblock available at
  \url{http://www.stat.berkeley.edu/~aldous/RWG/book.html}.

\bibitem[\protect\citename{Arenas {\em et~al.}\relax, }2008]{Arenas2008a}
Arenas, Alex, Díaz-Guilera, Albert, Kurths, J\"urgen, Moreno, Yamir, \& Zhou,
  Changsong. (2008).
\newblock {S}ynchronization in complex networks.
\newblock {\em Physics reports}, {\bf 469}(3), 93 -- 153.

\bibitem[\protect\citename{Barahona {\em et~al.}\relax, }1997]{Barahona1997}
Barahona, M., Tr\'ias, E., Orlando, T.~P., Duwel, A.~E., van~der Zant, H.
  S.~J., Watanabe, Shinya, \& Strogatz, S.~H. (1997).
\newblock {R}esonances of dynamical checkerboard states in {J}osephson arrays
  with self-inductance.
\newblock {\em Phys. rev. b}, {\bf 55}(May), R11989--R11992.

\bibitem[\protect\citename{Barahona \& Watanabe, }1998]{Barahona1998}
Barahona, Mauricio, \& Watanabe, Shinya. (1998).
\newblock {R}ow-switched states in two-dimensional underdamped
  {J}osephson-junction arrays.
\newblock {\em Phys. rev. b}, {\bf 57}(May), 10893--10912.

\bibitem[\protect\citename{Boccaletti {\em et~al.}\relax,
  }2006]{Boccaletti2006}
Boccaletti, S., Latora, V., Moreno, Y., Chavez, M., \& Hwang, D.-U. (2006).
\newblock {C}omplex networks: {S}tructure and dynamics.
\newblock {\em Physics reports}, {\bf 424}(4-5), 175 -- 308.

\bibitem[\protect\citename{Bonacich, }1987]{Bonacich1987}
Bonacich, Phillip. (1987).
\newblock {P}ower and {C}entrality: {A} {F}amily of {M}easures.
\newblock {\em American journal of sociology}, {\bf 92}(5), pp. 1170--1182.

\bibitem[\protect\citename{Borgatti, }2005]{Borgatti2005}
Borgatti, Stephen~P. (2005).
\newblock {C}entrality and network flow.
\newblock {\em Social networks}, {\bf 27}(1), 55 -- 71.

\bibitem[\protect\citename{Bozzo \& Franceschet, }2012]{Bozzo2012}
Bozzo, Enrico, \& Franceschet, Massimo. (2012).
\newblock {A}pproximations of the {G}eneralized {I}nverse of the {G}raph
  {L}aplacian {M}atrix.
\newblock {\em Internet mathematics}, {\bf 8}(4), 456--481.

\bibitem[\protect\citename{Brummitt {\em et~al.}\relax, }2012]{Brummitt2012}
Brummitt, Charles~D., D'Souza, Raissa~M., \& Leicht, E.~A. (2012).
\newblock {S}uppressing cascades of load in interdependent networks.
\newblock {\em Proceedings of the national academy of sciences}, {\bf 109}(12),
  E680--E689.

\bibitem[\protect\citename{Delmotte {\em et~al.}\relax, }2011]{Delmotte2011}
Delmotte, A, Tate, E~W, Yaliraki, S~N, \& Barahona, M. (2011).
\newblock {P}rotein multi-scale organization through graph partitioning and
  robustness analysis: application to the myosin–myosin light chain
  interaction.
\newblock {\em Physical biology}, {\bf 8}(5), 055010.

\bibitem[\protect\citename{Delvenne {\em et~al.}\relax, }2010]{Delvenne2010}
Delvenne, J.-C., Yaliraki, S.~N., \& Barahona, M. (2010).
\newblock {S}tability of graph communities across time scales.
\newblock {\em Proceedings of the national academy of sciences}, {\bf 107}(29),
  12755--12760.

\bibitem[\protect\citename{Delvenne \& Libert, }2011]{Delvenne2011}
Delvenne, Jean-Charles, \& Libert, Anne-Sophie. (2011).
\newblock {C}entrality measures and thermodynamic formalism for complex
  networks.
\newblock {\em Phys. rev. e}, {\bf 83}(Apr), 046117.

\bibitem[\protect\citename{Delvenne {\em et~al.}\relax, }2013]{Delvenne2012}
Delvenne, Jean-Charles, Schaub, Michael~T., Yaliraki, Sophia~N, \& Barahona,
  Mauricio. (2013).
\newblock {T}he stability of a graph partition: {A} dynamics-based framework
  for community detection.
\newblock  Ganguly, Niloy, Mukherjee, Animesh, Choudhury, Monojit, Peruani,
  Fernando, \& Mitra, Bivas (eds), {\em Time varying dynamical networks}.
\newblock Boston: Birkh\"auser, Springer.
\newblock to be published.

\bibitem[\protect\citename{Doyle \& Snell, }1984]{Doyle1984}
Doyle, P.G., \& Snell, J.L. (1984).
\newblock {\em {R}andom walks and electric networks}.
\newblock Carus mathematical monographs.
\newblock Mathematical Association of America.
\newblock open version available at
  \url{http://www.math.dartmouth.edu/~doyle/}.

\bibitem[\protect\citename{Evans \& Lambiotte, }2009]{Evans2009}
Evans, T.~S., \& Lambiotte, R. (2009).
\newblock {L}ine graphs, link partitions, and overlapping communities.
\newblock {\em Phys. rev. e}, {\bf 80}(1), 016105.

\bibitem[\protect\citename{Fortunato, }2010]{Fortunato2010}
Fortunato, Santo. (2010).
\newblock {C}ommunity detection in graphs.
\newblock {\em Physics reports}, {\bf 486}(3-5), 75 -- 174.

\bibitem[\protect\citename{Fouss {\em et~al.}\relax, }2007]{Fouss2007}
Fouss, Francois, Pirotte, Alain, Renders, Jean-Michel, \& Saerens, Marco.
  (2007).
\newblock {R}andom-{W}alk {C}omputation of {S}imilarities between {N}odes of a
  {G}raph with {A}pplication to {C}ollaborative {R}ecommendation.
\newblock {\em Knowledge and data engineering, ieee transactions on}, {\bf
  19}(3), 355 --369.

\bibitem[\protect\citename{Freeman, }1977]{Freeman1977}
Freeman, Linton~C. (1977).
\newblock {A} {S}et of {M}easures of {C}entrality {B}ased on {B}etweenness.
\newblock {\em Sociometry}, {\bf 40}(1), pp. 35--41.

\bibitem[\protect\citename{Freeman, }1978]{Freeman1978}
Freeman, Linton~C. (1978).
\newblock {C}entrality in social networks conceptual clarification.
\newblock {\em Social networks}, {\bf 1}(3), 215 -- 239.

\bibitem[\protect\citename{Ghosh {\em et~al.}\relax, }2008]{Ghosh2008}
Ghosh, Arpita, Boyd, Stephen, \& Saberi, Amin. (2008).
\newblock {M}inimizing {E}ffective {R}esistance of a {G}raph.
\newblock {\em Siam review}, {\bf 50}(1), 37--66.

\bibitem[\protect\citename{Godsil \& Royle, }2001]{Godsil2001}
Godsil, C.D., \& Royle, G.F. (2001).
\newblock {\em {A}lgebraic {G}raph {T}heory}.
\newblock Graduate Texts in Mathematics Series.
\newblock Springer New York.

\bibitem[\protect\citename{Guattery, }1998]{Guattery1998}
Guattery, Stephen. (1998).
\newblock {\em {G}raph {E}mbeddings, {S}ymmetric {R}eal {M}atrices, and
  {G}eneralized {I}nverses}.
\newblock Tech. rept. NASA/CR-1998-208462. Institute for Computer Applications
  in Science and Engineering NASA Langley Research Center.

\bibitem[\protect\citename{G\"{u}ler {\em et~al.}\relax, }2007]{Gueler2007}
G\"{u}ler, Teoman, Gross, George, \& Liu, Minghai. (2007).
\newblock {G}eneralized line outage distribution factors.
\newblock {\em Power systems, ieee transactions on}, {\bf 22}(2), 879--881.

\bibitem[\protect\citename{Harary \& Norman, }1960]{Harary1960}
Harary, Frank, \& Norman, Robert~Z. (1960).
\newblock {S}ome properties of line digraphs.
\newblock {\em Rendiconti del circolo matematico di palermo}, {\bf 9},
  161--168.

\bibitem[\protect\citename{Hardaker {\em et~al.}\relax, }2001]{Hardaker2001}
Hardaker, Laura~Anne, Singer, Emily, Kerr, Rex, Zhou, Guotong, \& Schafer,
  William~R. (2001).
\newblock {S}erotonin modulates locomotory behavior and coordinates egg-laying
  and movement in {C}aenorhabditis elegans.
\newblock {\em Journal of neurobiology}, {\bf 49}(4), 303--313.

\bibitem[\protect\citename{Hutchinson {\em et~al.}\relax,
  }1988]{Hutchinson1988}
Hutchinson, J., Koch, C., Luo, J., \& Mead, C. (1988).
\newblock {C}omputing motion using analog and binary resistive networks.
\newblock {\em Computer}, {\bf 21}(3), 52 --63.

\bibitem[\protect\citename{Jadbabaie {\em et~al.}\relax, }2004]{Jadbabaie2004}
Jadbabaie, A., Motee, N., \& Barahona, M. (2004).
\newblock {O}n the stability of the {K}uramoto model of coupled nonlinear
  oscillators.
\newblock {\em Pages  4296 --4301 of:} {\em Proceedings of the american control
  conference, 2004.},  vol. 5.

\bibitem[\protect\citename{Klein \& Randi\'c, }1993]{Klein1993}
Klein, D.~J., \& Randi\'c, M. (1993).
\newblock {R}esistance distance.
\newblock {\em Journal of mathematical chemistry}, {\bf 12}, 81--95.

\bibitem[\protect\citename{Lambiotte {\em et~al.}\relax, }2009]{Lambiotte2009}
Lambiotte, Renaud, Delvenne, Jean-Charles, \& Barahona, Mauricio. 2009 (Oct).
\newblock {\em {L}aplacian {D}ynamics and {M}ultiscale {M}odular {S}tructure in
  {N}etworks}.
\newblock arXiv:0812.1770.

\bibitem[\protect\citename{Lehmann \& Bernasconi, }2013]{Lehmann}
Lehmann, J., \& Bernasconi, J. (2013).
\newblock {in preparation}.

\bibitem[\protect\citename{Lehmann \& Bernasconi, }2010]{Lehmann2010}
Lehmann, J\"org, \& Bernasconi, Jakob. (2010).
\newblock {S}tochastic load-redistribution model for cascading failure
  propagation.
\newblock {\em Phys. rev. e}, {\bf 81}(Mar), 031129.

\bibitem[\protect\citename{Lov\'asz, }1994]{Lovasz1994}
Lov\'asz, L\'aszl\'o. 1994 (May).
\newblock {\em {R}andom walks on graphs - a survey}.
\newblock Tech. rept. YALEU/DCS/TR-1029. Yale University, Department Computer
  Science, New Haven CT 06520.

\bibitem[\protect\citename{Meyer~Jr., }1973]{Meyer1973}
Meyer~Jr., Carl~D. (1973).
\newblock {G}eneralized {I}nversion of {M}odified {M}atrices.
\newblock {\em Siam journal on applied mathematics}, {\bf 24}(3), pp. 315--323.

\bibitem[\protect\citename{Mohar, }1992]{Mohar1992}
Mohar, Bojan. (1992).
\newblock {L}aplace eigenvalues of graphs—a survey.
\newblock {\em Discrete mathematics}, {\bf 109}(1-3), 171 -- 183.

\bibitem[\protect\citename{Mohar \& Juvan, }1997]{Mohar1997}
Mohar, Bojan, \& Juvan, Martin. (1997).
\newblock {S}ome applications of {L}aplace eigenvalues of graphs.
\newblock {\em Pages  227--275 of:} {\em Graph symmetry: Algebraic methods and
  applications, volume 497 of nato asi series c},  vol. 497.

\bibitem[\protect\citename{Newman, }2005]{Newman2005}
Newman, M.E.~J. (2005).
\newblock {A} measure of betweenness centrality based on random walks.
\newblock {\em Social networks}, {\bf 27}(1), 39 -- 54.

\bibitem[\protect\citename{Poggio {\em et~al.}\relax, }1985]{Poggio1985}
Poggio, Tomaso, Torre, Vincent, \& Koch, Christof. (1985).
\newblock {C}omputational vision and regularization theory.
\newblock {\em Nature}, {\bf 317}(6035), 314--319.

\bibitem[\protect\citename{Rosas-Casals {\em et~al.}\relax,
  }2007]{Rosas-Casals2007}
Rosas-Casals, Mart\'{\i}, Valverde, Sergi, \& Sol{\'e}, Ricard~V. (2007).
\newblock {T}opological {V}ulnerability of the {E}uropean {P}ower {G}rid under
  {E}rrors and {A}ttacks.
\newblock {\em I. j. bifurcation and chaos}, {\bf 17}(7), 2465--2475.

\bibitem[\protect\citename{Saerens {\em et~al.}\relax, }2004]{Saerens2004}
Saerens, Marco, Fouss, Francois, Yen, Luh, \& Dupont, Pierre. (2004).
\newblock {T}he {P}rincipal {C}omponents {A}nalysis of a {G}raph, and {I}ts
  {R}elationships to {S}pectral {C}lustering.
\newblock {\em Pages  371--383 of:} Boulicaut, Jean-François, Esposito,
  Floriana, Giannotti, Fosca, \& Pedreschi, Dino (eds), {\em Machine learning:
  Ecml 2004}.
\newblock Lecture Notes in Computer Science, vol. 3201.
\newblock Springer Berlin / Heidelberg.

\bibitem[\protect\citename{Schaeffer, }2007]{Schaeffer2007}
Schaeffer, Satu~Elisa. (2007).
\newblock {G}raph clustering.
\newblock {\em Computer science review}, {\bf 1}(1), 27 -- 64.

\bibitem[\protect\citename{Schaub {\em et~al.}\relax, }2012a]{Schaub2012a}
Schaub, Michael~T., Lambiotte, Renaud, \& Barahona, Mauricio. (2012a).
\newblock {E}ncoding dynamics for multiscale community detection: {M}arkov time
  sweeping for the map equation.
\newblock {\em Phys. rev. e}, {\bf 86}(Aug), 026112.

\bibitem[\protect\citename{Schaub {\em et~al.}\relax, }2012b]{Schaub2012}
Schaub, Michael~T., Delvenne, Jean-Charles, Yaliraki, Sophia~N., \& Barahona,
  Mauricio. (2012b).
\newblock {M}arkov {D}ynamics as a {Z}ooming {L}ens for {M}ultiscale
  {C}ommunity {D}etection: {N}on {C}lique-{L}ike {C}ommunities and the
  {F}ield-of-{V}iew {L}imit.
\newblock {\em Plos one}, {\bf 7}(2), e32210.

\bibitem[\protect\citename{Sohn {\em et~al.}\relax, }2011]{Sohn2011}
Sohn, Yunkyu, Choi, Myung-Kyu, Ahn, Yong-Yeol, Lee, Junho, \& Jeong, Jaeseung.
  (2011).
\newblock {T}opological {C}luster {A}nalysis {R}eveals the {S}ystemic
  {O}rganization of the {C}aenorhabditis elegans {C}onnectome.
\newblock {\em Plos comput biol}, {\bf 7}(5), e1001139.

\bibitem[\protect\citename{Sol\'e {\em et~al.}\relax, }2008]{Sole2008}
Sol\'e, Ricard~V., Rosas-Casals, Mart\'{\i}, Corominas-Murtra, Bernat, \&
  Valverde, Sergi. (2008).
\newblock {R}obustness of the {E}uropean power grids under intentional attack.
\newblock {\em Phys. rev. e}, {\bf 77}(2), 026102.

\bibitem[\protect\citename{Spielman \& Srivastava, }2008]{Spielman2008}
Spielman, Daniel~A., \& Srivastava, Nikhil. (2008).
\newblock {G}raph sparsification by effective resistances.
\newblock {\em Pages  563--568 of:} {\em Proceedings of the 40th annual acm
  symposium on theory of computing}.
\newblock STOC '08.
\newblock New York, NY, USA: ACM.

\bibitem[\protect\citename{Strang, }1986]{Strang1986}
Strang, G. (1986).
\newblock {\em {I}ntroduction to {A}pplied {M}athematics}.
\newblock Wellesley-Cambridge Press.

\bibitem[\protect\citename{Varshney {\em et~al.}\relax, }2011]{Varshney2011}
Varshney, Lav~R., Chen, Beth~L., Paniagua, Eric, Hall, David~H., \& Chklovskii,
  Dmitri~B. (2011).
\newblock {S}tructural {P}roperties of the {C}aenorhabditis elegans {N}euronal
  {N}etwork.
\newblock {\em Plos comput biol}, {\bf 7}(2), e1001066.

\bibitem[\protect\citename{Watts \& Strogatz, }1998]{Watts1998}
Watts, Duncan~J., \& Strogatz, Steven~H. (1998).
\newblock {C}ollective dynamics of 'small-world' networks.
\newblock {\em Nature}, {\bf 393}(6684), 440--442.

\bibitem[\protect\citename{White {\em et~al.}\relax, }1986]{White1986}
White, J.~G., Southgate, E., Thomson, J.~N., \& Brenner, S. (1986).
\newblock {T}he {S}tructure of the {N}ervous {S}ystem of the {N}ematode
  {C}aenorhabditis elegans.
\newblock {\em Philosophical transactions of the royal society of london. b,
  biological sciences}, {\bf 314}(1165), 1--340.

\bibitem[\protect\citename{Witthaut \& Timme, }2012]{Witthaut2012}
Witthaut, Dirk, \& Timme, Marc. (2012).
\newblock {B}raess's paradox in oscillator networks, desynchronization and
  power outage.
\newblock {\em New journal of physics}, {\bf 14}(8), 083036.

\bibitem[\protect\citename{Wood \& Wollenberg, }1996]{Wood1996}
Wood, A.J., \& Wollenberg, B.F. (1996).
\newblock {\em {P}ower {G}eneration, {O}peration, and {C}ontrol}.
\newblock Wiley-Interscience, New York.

\bibitem[\protect\citename{Wu \& Huberman, }2004]{Wu2004}
Wu, F., \& Huberman, B.A. (2004).
\newblock {F}inding communities in linear time: a physics approach.
\newblock {\em The european physical journal b - condensed matter and complex
  systems}, {\bf 38}, 331--338.

\bibitem[\protect\citename{Youn {\em et~al.}\relax, }2008]{Youn2008}
Youn, Hyejin, Gastner, Michael~T., \& Jeong, Hawoong. (2008).
\newblock {P}rice of {A}narchy in {T}ransportation {N}etworks: {E}fficiency and
  {O}ptimality {C}ontrol.
\newblock {\em Phys. rev. lett.}, {\bf 101}(Sep), 128701.

\bibitem[\protect\citename{Yuan {\em et~al.}\relax, }2013]{Yuan2013}
Yuan, Y., Stan, G.-B., Shi, L., Barahona, M., \& Goncalves, J. (2013).
\newblock {D}ecentralised minimum-time consensus.
\newblock {\em Automatica}, {\bf 49}(5), 1227 -- 1235.

\end{thebibliography}

\end{document}